\documentstyle[12pt,aaspp4]{article}

\received{}
\accepted{}
\slugcomment{Accepted by the Astronomical Journal}

\begin{document}
\include{psfig}
\def\kms{km~s$^{-1}$ }
\def\Lya{Ly$\alpha$ }
\def\lya{Ly$\alpha$ }
\def\Lyb{Ly$\beta$ }
\def\lyb{Ly$\beta$ }
\def\Lyg{Ly$\gamma$ }
\def\Lyd{Ly$\delta$ }
\def\Lye{Ly$\epsilon$ }
\def\d{$d_5$ }
\def\Ly{Lyman}
\def\ang{\AA }
\def\gq{$\geq$ }
\def\zem{$z_{em}$ }
\def\zabs{$z_{abs}$ }
\def\cm2{cm$^{-2}$ }

%
\title{The Neutral Hydrogen Column Density towards Q1937--1009 from the
	Unabsorbed Intrinsic Continuum in the Lyman-$\alpha$ Forest}

\author{SCOTT BURLES\altaffilmark{1} \& DAVID TYTLER\altaffilmark{1}} 
\affil{Department of Physics, and Center for Astrophysics and Space
Sciences \\
University of California, San
Diego \\
C0424, La Jolla, CA 92093-0424}
 
\altaffiltext{1}{Visiting Astronomer, W. M. Keck Telescope, California
Association for Research in Astronomy}
 
%
%
 
\begin{abstract}

The absorption system
at $z=3.572$ towards Q1937--1009 provides the best extragalactic
measurement of the atomic deuterium to hydrogen ratio, D/H.
We have obtained a new low-resolution, high signal-to-noise ratio (SNR) 
Keck spectrum 
to re-measure the total neutral
hydrogen column density N(H~I) using the amount of Lyman continuum absorption.
We develop a new method to directly determine the
intrinsic 
unabsorbed quasar continuum from low-resolution spectra of the \Lya forest
for the first time.
We use three types of spectra to measure N(H~I): 
(1) A wide slit spectrum
for flux calibration, (2) a high-resolution spectrum to determine the
unabsorbed continuum between \Lya forest lines, and 
(3) a high SNR spectrum to measure the residual flux below the Lyman
limit.
We measure Log~[N(H~I)] = 17.86 $\pm$ 0.02 \cm2, which is reliable
because N(H~I) is fully specified by the data.
This result is consistent with the N(H~I) measured 
by Tytler, Fan \& Burles (1996) from the Lyman series absorption lines,
but not with absorption models proposed by Wampler (1996) 
nor estimates of the total N(H~I) by Songaila et al. (1997), both of which
suggested lower N(H~I) and higher D/H.
The Wampler models predict abundant flux below the Lyman limit which is
absent from both our old and new spectra, taken with different instruments.
The new Keck data is consistent with the data presented 
by Songaila et al. (1997).
The results differ due to the different methods of analysis,
and our measurement of the QSO continuum does not agree with the 
continuum models assumed by Songaila et al. (1997). 

\end{abstract}

\section{INTRODUCTION}

High-redshift QSO absorption systems can yield measurements
of the deuterium-to-hydrogen ratio (D/H) 
in nearly primordial gas (\cite{ada76}; \cite{tyt96}; \cite{bur96}).
Recently, limits on D/H have been presented
in the range of 2.3 $\times 10^{-5} <$ D/H $< 24 \times 10^{-5}$ 
(for a list, see Tytler \& Burles 1996).  But deuterium is rarely  
seen in absorption in QSO spectra.  
Utilizing low resolution spectra of over 300 QSOs
to identify systems
most likely to show deuterium (Burles et al. 1997),  
we have found only 2 systems
which will provide measures of D/H. 
No one else has verified a measurement.
Because of the small number of systems available, 
it is efficient to study each in as
much detail as possible.   
In this paper, we continue with our analysis of Q1937--1009.

Tytler, Fan \& Burles (1996, hereafter TFB) presented
high resolution spectra of Q1937--1009 and a
measurement of D/H in the absorption system at $z=3.572$.
This system is well suited for a measurement of D/H because the hydrogen
has a simple, narrow velocity structure, and the deuterium
feature is strong but unsaturated.
TFB measured the total neutral hydrogen column density, 
Log~[N(H~I)] $=17.94 \pm 0.06 \pm 0.05$ \cm2 (the first error is 
statistical, and the second is systematic), using 
a two component Voigt profile fit to 
the Lyman series absorption lines.

Wampler (1996) has suggested that this hydrogen column density is 
not well determined, due to either incorrect sky subtraction or
improper modeling.  He proposed models with 3--6 times less
N(H~I).  Lowering N(H~I) does two things: 
(1) it weakens the
absorption in the damping wings of \Lya, and  
(2) it raises the residual
flux below the Lyman limit at 4200 \AA. 
N(H~I) is difficult to measure from the dampings wings 
because systematic errors in the assumed unabsorbed QSO continuum level
are important. 
On the other hand, the total N(H~I) is well constrained by
the residual flux below the Lyman limit.  

Songaila et al. (1997) obtained LRIS spectra of Q1937--1009 to
determine the total N(H~I) at $z=3.572$ 
by measuring the residual Lyman limit flux.
The data which they present is consistent with the data
which we present below.  
But the final answers differ because we use different methods.
They report a confident upper limit, Log~[N(H~I)] $<$ 17.7 \cm2, which we
do not confirm, 
and our result is not consistent with 
theirs.  We discuss possible reasons for the differences in section 4.

We face two problems 
with this measurement, and both are due to the absorption
of the other unrelated lines in the 
\Lya forest.  
First, the high number density of the \Lya forest lines
makes it difficult to determine the intrinsic unabsorbed quasar continuum.
Second, \Lya absorption clouds at lower redshift contribute  
to the optical depth below the Lyman limit, which leads to
an overestimation of the optical depth at $z=3.572$.
 
We have obtained data on Q1937--1009 to overcome both these
difficulties.  We have three spectra, from three different
instruments and two different telescopes.
We refer to each spectrum by the instrument used:
LRIS, HIRES and Kast.  Each spectrum provides additional
and independent information to measure N(H~I) at $z = 3.572$.
The LRIS spectrum has high SNR at the wavelengths
below the Lyman limit.
The HIRES spectrum resolves all the \Lya forest
lines from \Lya emission at 5830 \AA~ to the Lyman limit at
4200 \AA, and allows us to measure the local continuum between the
unrelated \Lya forest
absorption lines.  The Kast spectrum will provide the absolute flux
of Q1937--1009, from the Lyman limit to 8000 \AA, which is needed to measure
the amount of Lyman continuum absorption.

\section{OBSERVATIONS \& DATA REDUCTION}

In this section, we present two new spectra of Q1937--1009: the Kast 
and LRIS spectra.  
We discuss the data reduction 
and the combination of these spectra with the
existing HIRES spectrum.  

\subsection{Kast Spectrum for Flux Calibration}

Q1937--1009 was observed on the night of August 19, 1996 with the Kast double
spectrograph on the Shane 3-m Telescope at Lick Observatory.
Two 3000 second exposures were obtained with
a 7" wide by 145" long slit and the 600 lines/mm
grating on both the red and blue sides.
The effective resolution of $\approx$ 2" is determined by 
the telescope guiding and seeing.
Flat field images were taken
immediately after the object exposures to remove the fringing
effects redward of 7000 \AA.  The flux standard BD +33$^o$ 2642 
was observed in the same setup directly before Q1937--1009.  
The spectrograph was rotated to align its slit
with the average parallactic angle in all observations.
The CCD images were overscan corrected, short bias subtracted,
and flat-field corrected.  
The spectrum was optimally extracted from these images in IRAF,
and placed on the HIRES wavelength scale (section 2.3). 
Extinction corrections for the Lick Observatory were applied to both 
the object and standard star spectra.
The spectrum was also dereddened with E(B-V) = 0.18 and R=3.2 
(\cite{bur84}).
Figure 1 shows the final flux calibrated spectrum of Q1937--1009.

\subsection{LRIS Spectrum}

On the night of August 11, 1996, Charles Steidel kindly 
obtained three 30 minute exposures 
with the Low Resolution Imaging Spectrograph (LRIS)
on the Keck1 10-m Telescope.  
The observations were made with
a 1.0" wide by 180" long slit and a 900 lines/mm grating blazed at 5500 \AA.
For each observation, the spectrograph was rotated to align its slit
with the parallactic angle
to reduce slit losses from differential atmospheric refraction. 
The detector was a Tektronix 2048x2048 pixel CCD with  24 $\mu$m square pixels.
At the Cassegrain focus, each pixel sampled 0.22" square.  
The CCD images were overscan corrected, short bias subtracted,
and flat-field corrected.  
Each image was background subtracted by fitting
a 2nd order Chebyshev polynomial to each column.  Fifty pixels along the
spatial direction in an 11" region 
on each side of the object were used in the fits.  The three images
were then shifted spatially to overlay the object.   
The three images were coadded
and the 1-D spectrum was extracted from this final image.   Each pixel
in the 1-D spectrum is the sum of pixels in a 6.6" synthetic 
aperture centered on the
object.  The aperture center was traced along the object and fit 
with a 2nd order cubic spline.  The 1$\sigma$ error on each pixel includes
Poisson errors from the object and background and the readout noise of
the CCD.  We applied the HIRES wavelength scale described in section 2.3.

An initial flux calibration was accomplished with an extinction corrected
spectrum of BD +33$^o$ 2642.  We then
corrected the LRIS flux to match the absolute
flux of the Kast spectrum.  We smoothed the LRIS spectrum to the Kast
spectral resolution, and divided the smoothed LRIS spectrum into the Kast
spectrum.  A low order polynomial was fit to this response function and
applied to the original LRIS spectrum.  The response function removes the
low-order flux differences between the wide slit and narrow slit spectra.
The spectra were overlaid and the overall flux match between the LRIS
and Kast spectra were confirmed.
The fully reduced LRIS spectrum is shown in Figure 2.

\subsection{Common Wavelength Scale}

Our analysis requires nearly
identical wavelength scales on all spectra, hence we transferred
the wavelength scale from the existing HIRES spectrum (TFB) 
to the low resolution LRIS and Kast spectra.
We performed the following procedure for the LRIS spectrum.
We used a similar procedure to wavelength calibrate the Kast spectrum.

The HIRES spectrum 
(FWHM = 9 \kms) was smoothed to the LRIS spectral resolution (FWHM = 3.0 \AA)
and rebinned to the LRIS pixel size 
($\Delta \lambda$ = 0.83 \AA) and pixel positions.  We refer to this
spectrum as the Smoothed HIRES spectrum.
Unblended Lyman-$\alpha$ absorption
features were identified in the Smoothed HIRES spectrum and were
marked in the LRIS spectrum.  A 3rd order cubic spline was fit to the positions
of the absorption features in the LRIS spectrum,
and left an RMS residual of 0.22 pixels.
This procedure applied a high resolution wavelength calibration to
a low resolution spectrum.  

\section{The Intrinsic Unabsorbed Quasar Continuum}

In the LRIS spectrum, the 
intrinsic unabsorbed quasar continuum level is difficult to determine blueward
of Lyman-$\alpha$ emission, because blended \Lya features reduce the
observed flux to well below the unabsorbed quasar continuum in 
nearly all pixels.
We avoid this problem by transferring
the continuum level from the HIRES spectrum to the LRIS spectrum.
The HIRES spectrum resolves the \Lya lines and allows an unbiased estimate
of the continuum level in each echelle order (eg. \cite{kir97}).

We remove the absorption features in the LRIS
spectrum by dividing by the Smoothed HIRES spectrum.  
Both spectra have very high
SNR, and the division produces a fairly well behaved quotient 
which we show in Figure 3.   The result represents the intrinsic unabsorbed
continuum of a high redshift QSO, seen for the first time.
The break in the continuum near 4650\AA~ (1020 \AA~rest)  
agrees with the power-law break seen
in composite spectra of low-redshift quasars (\cite{zhe97}).
Lu \& Zuo (1994) showed that the unabsorbed continuum
in the \Lya forest deviates systematically from the extrapolated continuum.
This method does not require a statistical modeling of the absorption
spectrum or the extrapolation of the continuum from wavelengths redward
of \Lya emission.  

\subsection{Uncertainty in the Unabsorbed Continuum Level}

In Figure 3, the points represent the calculated continuum level
in each LRIS pixel.
The strong feature at 4920 \AA~ is the QSO \Lyb-O~VI emission 
feature.
The position of \Lyg is also marked at 4680 \AA.
We fit the quasar continuum with least squares minimization
and a linear fit, I($\lambda$) = a ($\lambda-$ 4550\AA) + b,
between 4250 \AA~ and 4850 \AA.
The best fit is the solid line in Figure 3, with $a=0.0028 \pm 0.0002$ and
$b=7.76 \pm 0.03$.
The dashed lines represent the 1$\sigma$ errors in the linear coefficients.
The linear model gives a good fit to the data, and provides a simple
extrapolation to lower wavelengths.

The scatter in the data is approximately
twice as large as expected for purely random
errors.
There are two probable reasons for the increased scatter:
(1) intrinsic features in the unabsorbed quasar spectrum, and
(2) systematic differences between the LRIS and Smoothed HIRES spectra.
Any residual differences in the wavelength scales in the two spectra
will produce greater than random scatter in the resulting spectrum.  
The scatter does increase towards lower wavelengths, 
and this is most likely due to
decreasing SNR with decreasing wavelengths, and intrinsic features in
the QSO spectrum.
Our guess at the 
systematic error in the continuum is 
shown as the dotted lines in Figure 3.
Assuming the unabsorbed continuum can be modeled as a low order polynomial,
the slope of the continuum in the region of interest is much better
determined than the absolute level.  
Even with these conservative estimates, the continuum is known to better
than 10\% over the
full spectral range in Figure 3.
We extrapolate the fit from 4200 \AA~ to 3850 \AA, which is the region where
the flux level drops due to the Lyman limit at $z=3.572$.  

With large optical depths, N(H~I) is fairly 
insensitive to the placement of the continuum.
A small error in the continuum, $\Delta I_0$, would result in 
a proportional error in the column density of
\begin{equation}
{{\Delta \rm{N(H~I)}} \over {\rm{N(H~I)}}} = {{\Delta I_0} \over {\tau}},
\end{equation}
which is reduced by the optical
depth of the absorption.  In our case, a 10\% error in the continuum level 
($\Delta I_0 = 0.1$)
for an optical depth of $\tau_c$ = 4.6, contributes an additional 2\% 
uncertainty in N(H~I).  This error is approximately one-half the magnitude of 
the statistical error discussed in the next section.

\section{N(H~I) Column Density}

The total neutral hydrogen column density, N(H~I), can be measured from
the Lyman continuum optical depth, $\tau(\lambda)$ = $-$Ln$(I_{\lambda}/I_0)$,
where $I_0$ is the unabsorbed flux, and $I_{\lambda}$ is the absorbed flux 
blueward of the Lyman limit at $\lambda_0 = 911.75 (1+z_{abs})$ \AA.
The hydrogen photoionization cross-section has an energy dependence 
which gives an optical depth as a function of wavelength of
\begin{equation}
\tau(\lambda) \approx {{\rm{N(H~I)}} 
\over {1.6\times10^{17} \rm{cm}^{-2}}} \left({\lambda \over
	{\lambda_0}}\right)^{2.7} 
\end{equation}
(\cite{ost89}).

If the absorption system at $z=3.572$ was the only one with significant
absorption at $\lambda < 4200$ \AA, we could fit eqn. 2 to the LRIS 
spectrum.  However we must first account for 
extra absorption, which can be placed into two groups, (1) the
absorption of higher-order Lyman lines (\Lyb, \Lyg, \Lyd, ...) and Lyman
continuum from high redshift systems
($2.75 < z <3.572$), and (2) Lyman-$\alpha$ absorption falling directly
in the region of interest ($2.16 < z < 2.45$).  
Absorption by hydrogen systems at higher redshifts ($3.572 < z < 3.78$)
has already been accounted for in the calculation of the quasar
continuum, because all their Lyman series lines and Lyman continuum absorption
lie at wavelengths greater than 4200 \AA, where the HIRES spectrum has
high SNR.
The group (1) absorption 
can be directly identified
from the \Lya lines in regions of the HIRES spectrum where the SNR is high.  
We fit Voigt profiles to measure the absorption parameters
for all systems in group (1) (Burles 1997).  
The corresponding \Lyb, \Lyg, \Lyd, ..., and Lyman 
continuum absorption is shown in Figure 4.

Absorption from group (2) cannot be directly measured, because 
the SNR is below one in the HIRES spectrum blueward of the Lyman limit.
We used Monte Carlo methods 
to statistically account for this absorption, 
and to determine simultaneously
the likelihood of the data as a function of the 
N(H~I) at $z=3.572$. 
We step through a range of N(H~I) values, and generate 10000 
model spectra for each value of N(H~I).
All models had the same absorption from the systems in group 1, but each
had a different realization of the group (2) \Lya lines, drawn from 
the known distribution of \Lya lines (\cite{kir97}).
Each model spectrum was compared to the data, and the probability
of each model was calculated with the $1\sigma$ error array.
The resulting likelihood function is shown in Figure 5.
The most likely value for the total N(H~I) is Log~[N(H~I)] = 17.86 \cm2.
Near the maximum likelihood, the function is nearly Gaussian and
gives a 1$\sigma$ error of 0.015 in logarithmic units.
Figure 4 shows the residual flux below 4200 \AA, and the model fit
with N(H~I) = 17.86 \cm2 plus the additional absorpition from group 1, 
but not the absorption from group 2.  
Adding the uncertainty from the continuum level to the 
likelihood function in Figure 5, 
we find Log~[N(H~I)] $= 17.86 \pm 0.02$ \cm2.

\section{Comparison of N(H~I) Measurements}

TFB measured the line profiles
of all uncontaminated Lyman lines and found Log~[N(H~I)] = 17.94 
$\pm$ 0.05 $\pm$ 0.06 \cm2, where the first error is statistical and
the second is systematic.  Now using different instruments and methods, 
we find an independent measure of Log~[N(H~I)] = 17.86 $\pm 0.02$, which agrees.

Songalia et al. (1997) used a similar LRIS spectrum, 
and reported a ``reasonable maximum", Log~[N(H~I)] $<$ 17.7 \cm2.
The inconsistency between the two results arises from the
different methods of analysis.  We present four reasons for this
inconsistency. 
(1) Lacking high-resolution spectra,
they relied on statistical models
to determine the unabsorbed quasar continuum.  
They proposed two 
models, both of which underestimate the unabsorbed continuum which 
we have measured. 
In their Figure 2, their model continua 
pass below their data, which is unphysical.
(2) They do not include most of the absorption systems in groups 1 or 2.
(3) Instead of using all the $\approx$ 350 pixels
in the Lyman limit region, 
they measure $\tau$ from a region with less than 20 pixels.
(4) They did not include the energy dependence in the hydrogen 
photoionization cross-section, which gives a 7\% underestimate of N(H~I). 

Our LRIS spectrum is apparently consistent with theirs.
A direct comparison of the data is difficult, 
because they do not
present their data with absolute flux units.
We transformed our fluxed LRIS
spectrum into Log Flux in units of energy per unit frequency.  We see
the same features, the same slope in flux, and the same decrease below the
Lyman limit.  In Figure 6, we show our LRIS spectrum with the same
relative scale and the same spectral resolution as their Figure 2a 
spectrum.
We measure the average flux blueward of 4200 \AA~ at Log Flux = 0.7, 
which appears to be consistent with their LRIS spectrum.  

Our result is not consistent with the alternative models proposed 
by Wampler (1996).   
His models require large systematic errors in the sky subtraction of
the HIRES spectrum presented by TFB.  The models do not agree with the data
for the folowing reasons:
The spectrum was optimally extracted using
automated routines written by Tom Barlow.  All optically thick Lyman Lines
are black at their line centers in the HIRES spectrum, which is a good check
on the sky subtraction.  In addition, 
two LRIS spectra (Songaila's and ours) both show
the same flux below the Lyman limit, and giving an N(H~I) which
is 50\% too large for Wampler's models.

We directly determined the intrinsic unabsorbed quasar continuum
using three independent spectra, and measured the 
optical depth of the Lyman limit at $z=3.572$.
We find Log~[N(H~I)] = 17.86 $\pm 0.02$, where the error includes both
photon noise and uncertainty in the continuum.
The result is consistent with that which we determined from fitting the
Lyamn series lines in the HIRES spectrum (TFB).
With this new constraint, we can now provide a better measurement of
D/H in this absorption system.
The N(D~I) reported by TFB cannot be
used because the deuterium and hydrogen profiles are blended (i.e. changing
N(H~I) will change N(D~I)).
Our improved analysis of 
the high-resolution spectra with this
new constraint will be presented in in Burles \& Tytler (1997).

\acknowledgements

We are very grateful to Charles Steidel for obtaining the LRIS observations.  
We thank Tom Barlow, David Kirkman, and Jason X. Prochaska for many helpful
suggestions.

\clearpage

\begin{figure}
\figurenum{1}
\centerline{
\psfig{figure=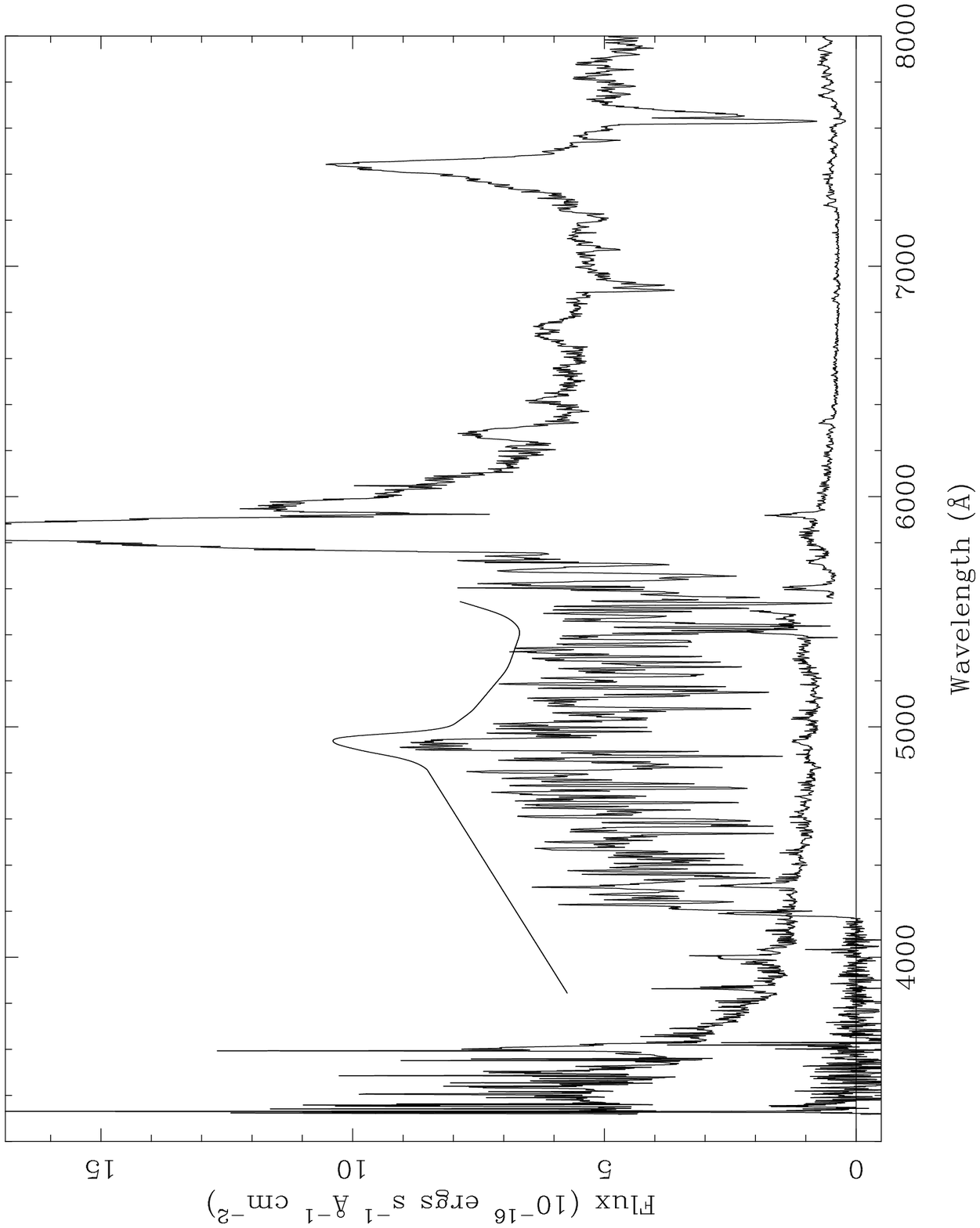,height=7.8in}}
\caption{ A wide slit spectrum of Q1937--1009 (\zem=3.805, V=17.5)
obtained at the Lick Observatory Shane 3-m telescope
with the Kast spectrograph.  This spectrum was used to flux
calibrate the LRIS spectrum. 
We show the unabsorbed quasar continuum (the smooth solid line) 
for reference.
The 10$\sigma$ error is also plotted, which is $\simeq 1.5$ at 4000 \AA. }
\end{figure}

\begin{figure}
\figurenum{2}
\centerline{
\psfig{figure=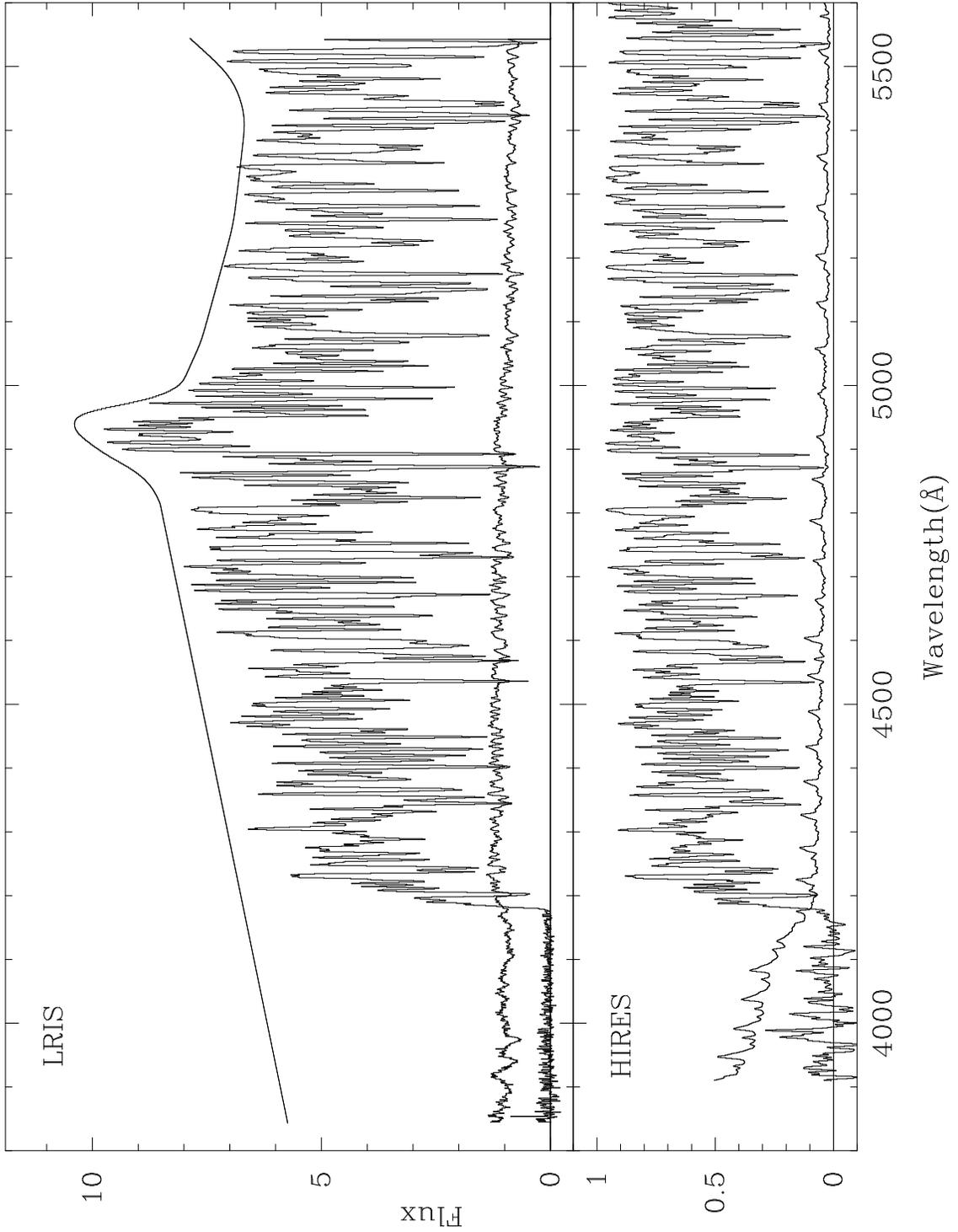,height=7.3in}}
\caption{
The top panel shows the spectrum of Q1937--1009
obtained at the 10-m W.M. Keck 1 Telescope with the Low Resolution Imaging
Spectrograph (LRIS).  The flux is in units of 10$^{-16}$ ergs s$^{-1}$ cm$^{-2}$
\AA$^{-1}$.  We also show the accompanying 10$\sigma$ error, which is $\simeq 1.0$ at 4000 \AA. 
The bottom panel
shows the HIRES data rebinned to LRIS pixels and smoothed to the LRIS resolution
(3\AA~ FWHM).  The HIRES
spectrum has been normalized with the unabsorbed quasar continuum set to 1.
We show the weighted 10$\sigma$ error per LRIS pixel, which is $\simeq 0.35$ at 4000 \AA.  }
\end{figure}

\begin{figure}
\figurenum{3}
\centerline{
\psfig{figure=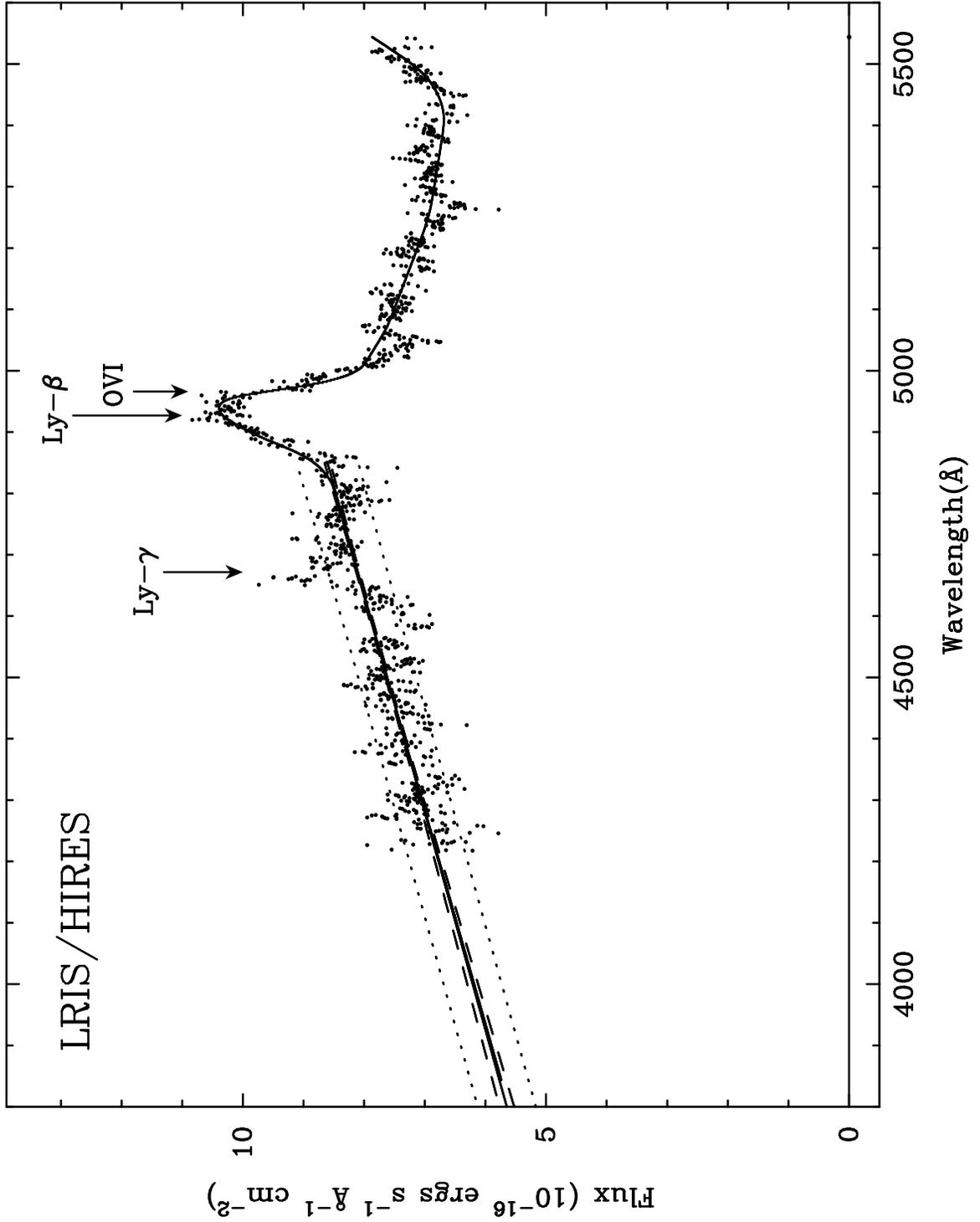,height=7.8in}}
\caption{
The LRIS spectrum divided by the HIRES spectrum.  
The points are pixels, and the smooth solid line is the adopted 
unabsorbed continuum.
The dashed lines show the 1$\sigma$ uncertainties in the linear fit to
the continuum below 4850 \AA.  Dotted lines indicate guessed systematic errors.}
\end{figure}

\begin{figure}
\figurenum{4}
\centerline{
\psfig{figure=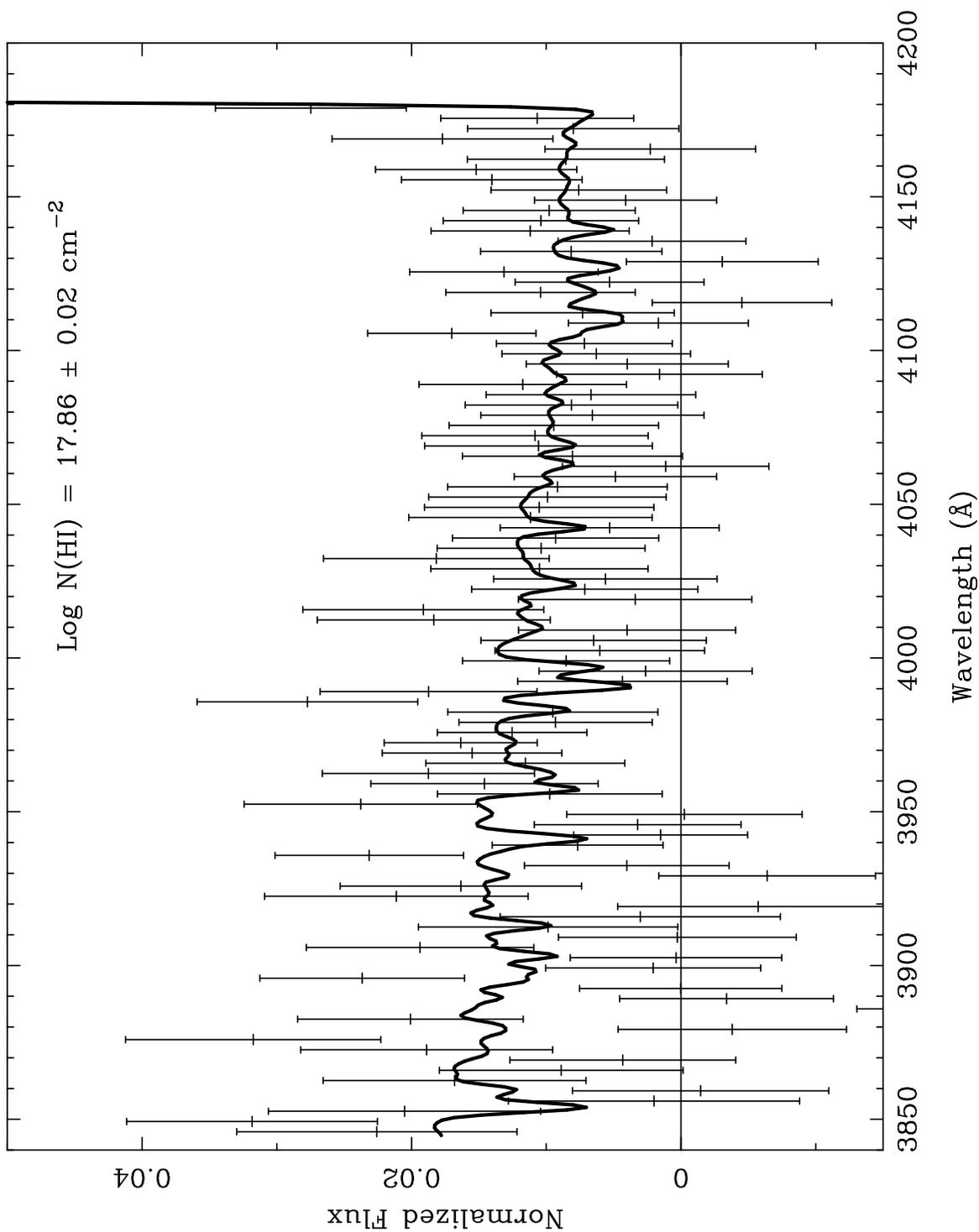,height=7.8in}}
\caption{
The portion of the LRIS spectrum showing residual flux below
the Lyman limit at $z=3.572$.  The data is shown as normalized flux
with 1$\sigma$ error bars, and is binned over 4 pixels for display purposes
only.
The fit of maximum likelihood is the solid curve corresponding to
Log~[N(H~I)] = 17.86 \cm2.}
\end{figure}

\begin{figure}
\figurenum{5}
\centerline{
\psfig{figure=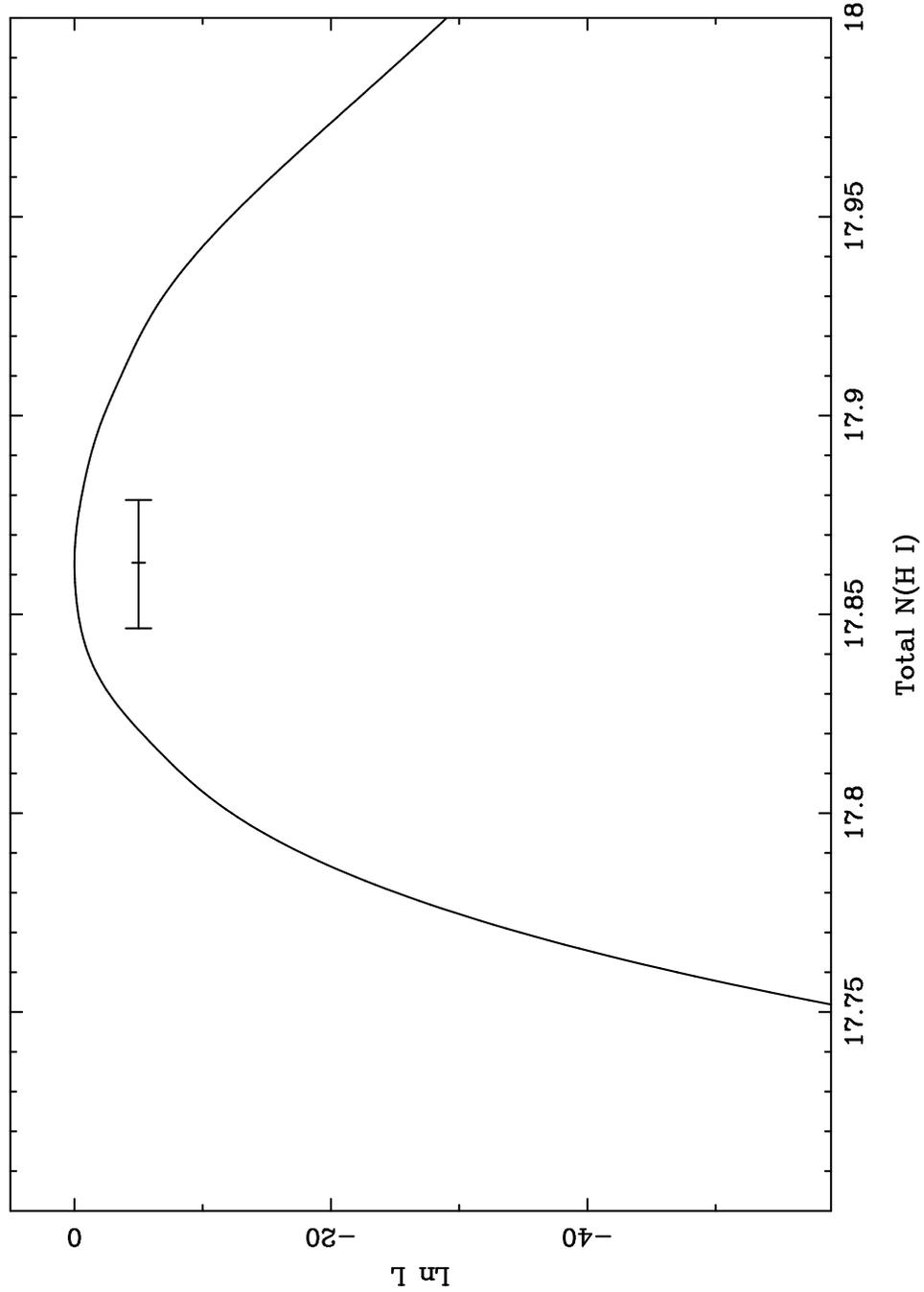,height=7.8in}}
\caption{
Likelihood function of N(H~I) from simulations of the LRIS data.
The maximum likelihood is at Log~[N(H~I)] = 17.863 \cm2, with a 67\% confidence
levels at Log~[N(H~I)] = 17.848 \cm2 and 17.878 \cm2.}
\end{figure}

\begin{figure}
\figurenum{6}
\centerline{
\psfig{figure=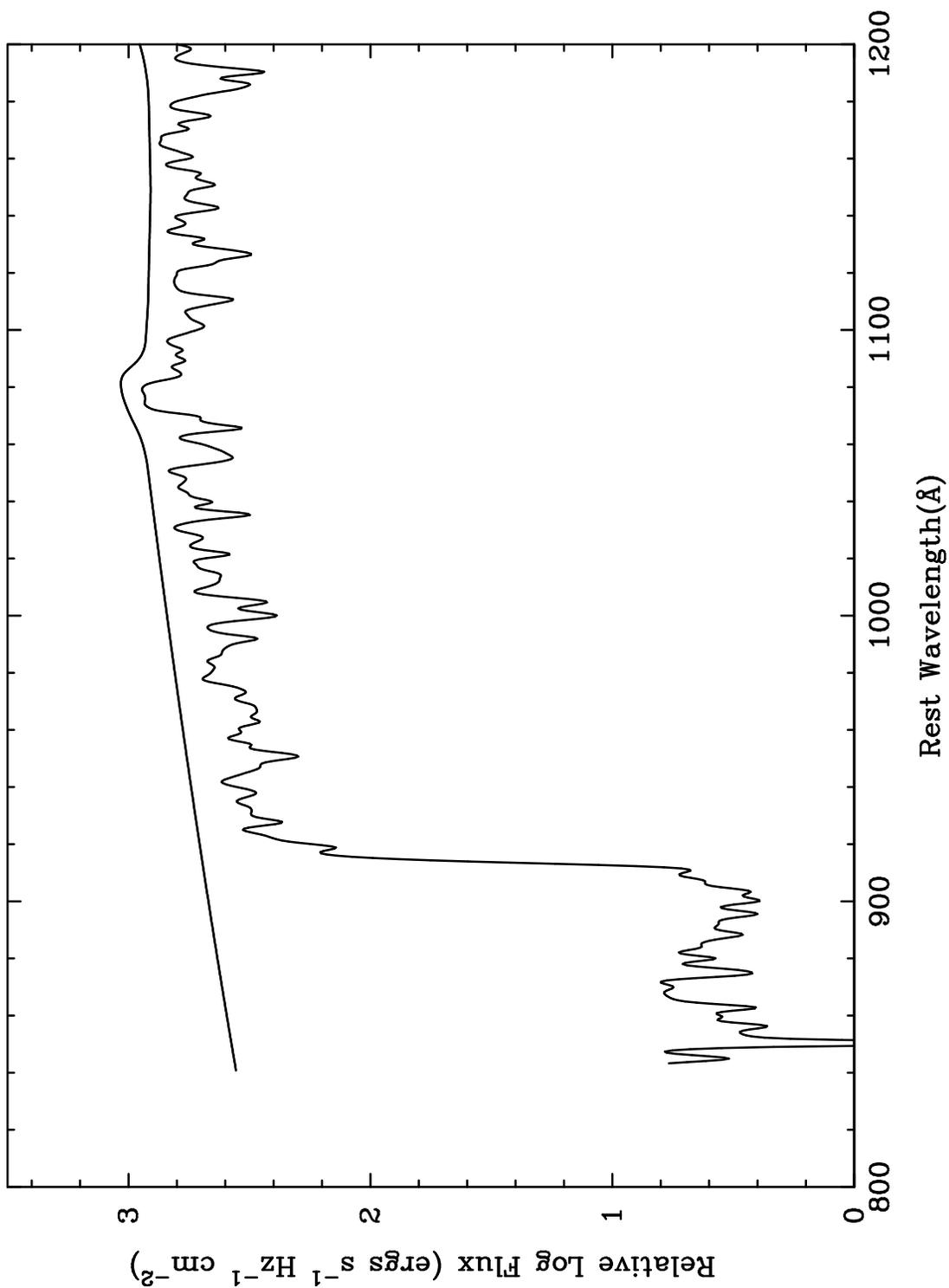,height=7.8in}}
\caption{
The LRIS spectrum shown in Figure 2 plotted in Relative Log
Flux in energy per unit frequency  
versus rest wavelength at $z=3.572$.  The smooth line is the 
intrinsic unabsorbed quasar continuum shown in Figure 3.
This figure is intended to facilitate
a comparison of our LRIS data with that presented by Songaila et al. (1997)
in their Figure 2a.}
\end{figure}

\end{document}